\documentclass[aps,twocolumn,showpacs]{revtex4}
\usepackage{graphicx}
\usepackage{epsfig}
\usepackage{amsfonts}
\usepackage{amsmath,amssymb}
\begin{document}
\title{Phase Diagram of the Proton-Neutron Interacting Boson Model}
\author{J.M. Arias$^1$,  J.E. Garc\'{\i}a--Ramos$^2$, and J. Dukelsky$^3$}
\affiliation {$^1$ Departamento de F\'{\i}sica At\'omica, Molecular y Nuclear,
Facultad de F\'{\i}sica, \\ Universidad de Sevilla, Apartado~1065,
41080 Sevilla, Spain.\\
$^2$ Departamento de F\'{\i}sica Aplicada, Universidad de Huelva,
21071 Huelva, Spain.\\
$^3$ Instituto de Estructura de la Materia, CSIC, Serrano 123,
28006 Madrid, Spain. }


\begin{abstract}
{We study the phase diagram of the proton--neutron interacting boson
model (IBM--2) with special emphasis on the phase transitions leading
to triaxial phases. The existence of a new critical point between
spherical and triaxial shapes is reported.}
\end{abstract}
\maketitle

Quantum phase transitions (QPT) have become a subject of great
interest in the study of several quantum many--body systems in
condensed matter, quantum optics, ultra--cold quantum gases, and
nuclear physics. QPT are structural changes taking place at zero
temperature as a function of a control parameter (for a recent review
see \cite{Vojta}). Examples of control parameters are the magnetic
field in spin systems, quantum Hall systems, and ultra-cold gases
close to a Feshbach resonance, or the hole-doping in cuprate
superconductors.

The atomic nucleus is a finite system composed of N neutrons and Z
protons (Z+N$\approx $100).
Though strictly speaking QPT take place for large systems in
the thermodynamic limit, finite nuclei can show the precursors of a
phase transition for some particular values of N and Z. In these
cases, one finds specific patterns in the low
energy spectrum revealing the strong quantum fluctuations responsible
for the phase transition \cite{IZ04}.
Recently the concept of critical point symmetry has been proposed by
Iachello and applied to atomic nuclei.
First, the transition from spherical to deformed
$\gamma$-unstable shapes was studied and the corresponding critical
point called E(5) \cite{E5}. Since then, the interest in
nuclear shape--phase transitions has been constantly growing.
The characteristics of the critical point in the phase transition from
spherical to axially deformed nuclei, called X(5), were presented in
Ref. \cite{X5}.
More recently, the critical point in the phase transition from axially
deformed to triaxial nuclei, called Y(5), has been analysed
\cite{Y5}. In all these cases, critical points are defined in the
context of the collective
Bohr Hamiltonian \cite{BMII}. Using some simplifying approximations
precise parameter--free predictions for several observables are
obtained. This allows to identify nuclei at the critical points
looking at spectroscopic properties. Indeed, some experimental candidates to
critical nuclei have already been proposed \cite{E5Exp,X5exp}.

The collective Bohr Hamiltonian, underlying this approach to critical
point symmetries, is closely related to the Interacting Boson Model
(IBM) \cite{IBM}. The simplest version of the IBM is called IBM--1
since in it no explicit distinction is made between protons and
neutrons. In IBM--1 there are three dynamical symmetries:
$SU(5)$, $O(6)$, and $SU(3)$. These
correspond to well defined nuclear shapes: spherical,
deformed $\gamma$-unstable, and prolate axial deformed,
respectively. The structure of the IBM--1 Hamiltonian allows to study
systematically the transition from one shape to another.
There were some pioneering works along these lines in the 80's
\cite{DSI80,FGD,Alex},
but it has been the recent introduction of the concept of critical point
symmetry that has recalled the attention of the community to the topic
of quantum phase transitions in nuclei. The phase diagram of the
IBM--1 has been studied from several points of view
\cite{DSI80,FGD,Alex,Cas,Jo1,Ar3}. The three different phases are
separated by lines of first order phase transition, with a singular
point in the transition from spherical to deformed $\gamma$-unstable
shape that is second order. In the usual IBM--1 no triaxial shapes
appear. These can only be stabilised
with the inclusion of specific three body forces. A more natural way
to generate triaxial deformations is by explicitly taking into account
the proton-neutron degree of freedom with the more realistic IBM--2
\cite{IBM-2}.

In this letter we will study the phase diagram of the IBM--2 using a
simplified Hamiltonian that keeps all the main ingredients of the most
general one. This is the Consistent--Q IBM--2 Hamiltonian \cite{Ca}
\begin{equation}
H= x \left(n_{d_\pi}+n_{d_\nu} \right)-\frac{1-x}{N}
Q^{(\chi_\pi,\chi_\nu ) }\cdot Q^{(\chi_\pi,\chi_\nu )} ~,
\label{HQ}
\end{equation}
where $n_{d}=\sum_{\mu }d_{\mu }^{\dagger }d_{\mu }$,
$Q^{(\chi_\pi,\chi_\nu )}=\left(Q_\pi^{\chi_\pi}+
Q_\nu^{\chi_\nu}\right)$ with
$Q^\chi_{\kappa }=\left[ d_{\kappa }^{\dagger }\widetilde s_{\kappa }+
s_{\kappa }^{\dagger }\widetilde d_{\kappa }\right]^{2} +
\chi_\kappa \left[ d_{\kappa }^{\dagger }\widetilde{d}_{\kappa }\right]^{2}$
and $N$ is the total number of bosons, which is equal to the number of
valence proton plus neutron pairs. The IBM phase diagram studied up to
now corresponds to the selection $\chi_\pi=\chi_\nu$ which produces
either spherical, axial or $\gamma$ independent shapes. We will extend
the previous works on IBM phase transitions by exploring the
transitions from axial to triaxial shapes within the mean field or
intrinsic state formalism. The trial wave function is the most general
proton-neutron boson condensate \cite{GK80,BM80,GLnp},
$|g\rangle=|N_\pi, N_\nu, \beta_\pi, \gamma_\pi , \beta_\nu,
\gamma_\nu, \Omega \rangle$
\begin{equation}
|g \rangle = \frac{(\Gamma^\dag_\pi) ^{N_\pi}
\hat{R}_3(\Omega)(\Gamma^\dag_\nu )^{N_\nu}}{\sqrt{N_\pi! N_\nu!}}
|0 \rangle \label{STIN}
\end{equation}
with
\begin{eqnarray}
\Gamma^\dagger_\kappa=\frac{1}{\sqrt{1+ \beta_\kappa^2}}
\left[s_\kappa^\dagger \right.&+& \beta_\kappa \cos \gamma_\kappa
d^\dagger_{\kappa 0}
  \nonumber \\
&+& \left. \frac{1}{\sqrt{2}} ~ \beta_\kappa
\sin \gamma_\kappa (d^\dagger_{\kappa 2} + d^\dagger_{\kappa -2}) \right]
\end{eqnarray}
where $\kappa=\pi,\nu$ and $\hat{R}_3(\Omega)$ is the three dimensional
rotation operator with $\Omega$ fixing the relative orientation (Euler
angles) between the proton and neutron condensates.
$N_\pi$ and $N_\nu$ are the numbers of valence proton and
neutron pairs, respectively. The equilibrium values of the structure
parameters ($\beta_\pi, \gamma_\pi , \beta_\nu, \gamma_\nu, \Omega$)
and the energy of the system for given values of the control
parameters in the Hamiltonian ($x, \chi_\pi, \chi_\nu$) can be
obtained by minimising the expectation value of the Hamiltonian
(\ref{HQ}) in the intrinsic state (\ref{STIN}):
$\delta \langle g |H|g\rangle = 0$.
\begin{figure}[hbt]
\includegraphics[height=6cm,width=6.5cm]{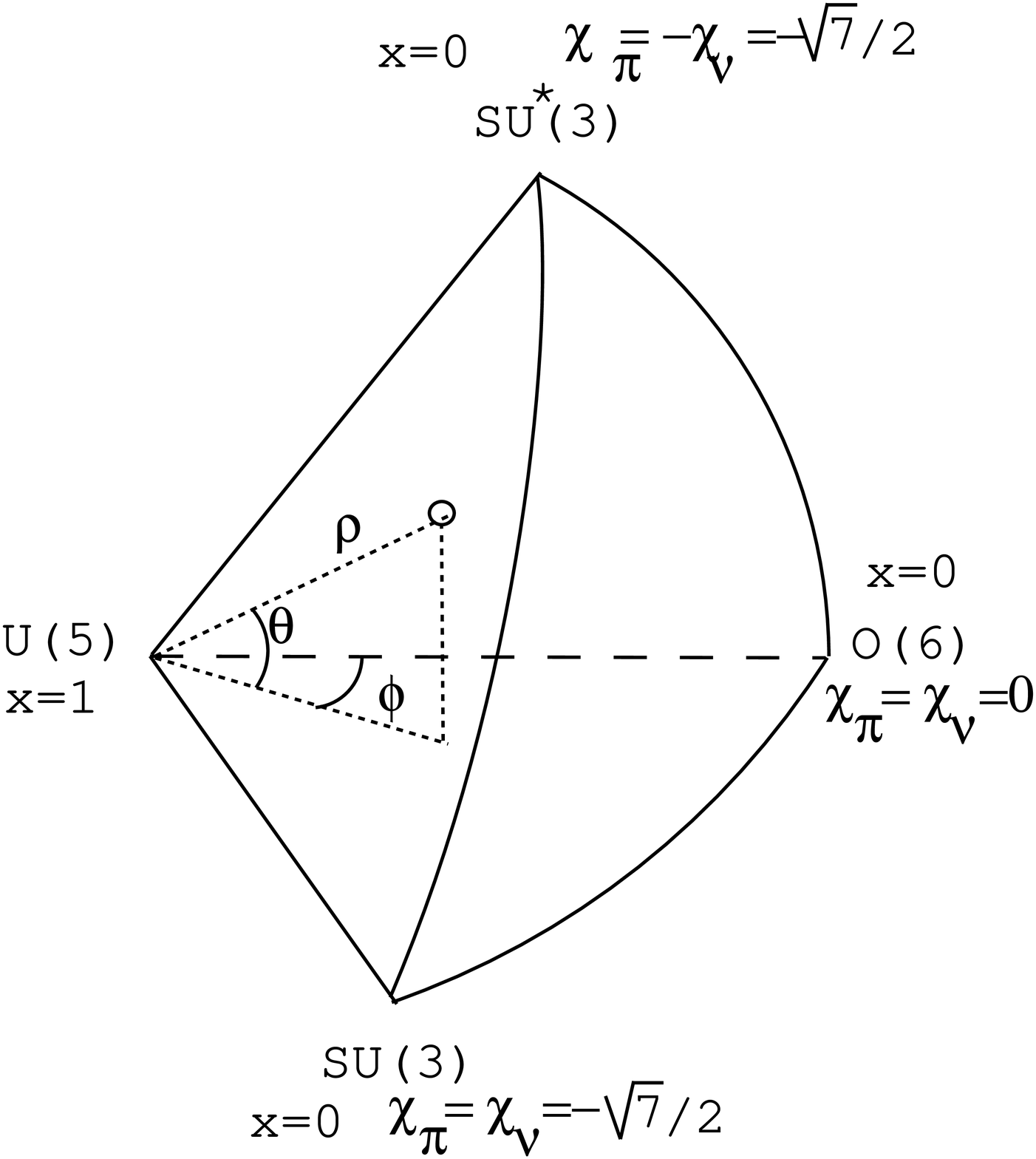}
\caption{Pictorial  representation of the IBM-2 parameter space
with a dynamical symmetry in each of the four vertices.} \label{FIG1}
\end{figure}
Although there is an explicit dependence of the energy on the Euler
 angles, it has been shown
\cite{GLnp} that oblique configurations (relative orientation angles
different from the aligned $\Omega=0$ or the perpendicular
$\Omega=\pi/2$ ) require a repulsive hexadecapole $\pi-\nu$
interaction. Therefore, since our Hamiltonian (\ref{HQ}) has no
hexadecapole terms, we do not expect oblique configurations.  We can
then safely assume that any arbitrary local minimum will have
$\Omega=0,~\gamma_\pi=\gamma_\nu =0^\circ $ (or equal to $60^\circ$)
for the aligned configurations or
$\Omega=0,~\gamma_\pi =0^\circ,~\gamma_\nu =60^\circ$ (or
$\gamma_\pi =60^\circ,~\gamma_\nu =0^\circ$)
for the perpendicular
configurations. In both cases $\Omega=0$ and the rotation operator
disappears from the intrinsic state (\ref{STIN}).
In that situation, the energy per boson in the limit $N_\pi,N_\nu
 \rightarrow \infty $  reduces to
\begin{eqnarray}
&&E(\beta_\pi, \gamma_\pi , \beta_\nu, \gamma_\nu;\chi_\pi, \chi_\nu, x)= x
\sum_{\kappa=\pi,\nu} \frac{\beta_\kappa^2}{1+\beta_\kappa^2} \nonumber \\
&-& \frac{1-x}{4} \sum_{\mu=0,\pm 2} \left[\sum_{\kappa=\pi,\nu}
  Q_\mu^2(\kappa) +  2 Q_\mu(\pi)  Q_{-\mu}(\nu) \right]
\end{eqnarray}
where we have used the notation
$Q_0(\kappa)=Q_0(\beta,\gamma,\chi)_\kappa$ and $Q_2(\kappa)=Q_{-2}(\kappa)=
Q_2(\beta,\gamma,\chi)_\kappa=Q_{-2}(\beta,\gamma,\chi)_\kappa$ with
\begin{eqnarray}
Q_0(\kappa)&=&\frac{1}{1+\beta_\kappa^2}{\left[ 2 \beta_\kappa
  \cos \gamma_\kappa   - \sqrt{\frac{2}{7}}
\beta_\kappa^2 \chi_\kappa \cos (2\gamma_\kappa) \right]}~,
\nonumber\\
Q_2(\kappa)&=&\frac{1}{1+\beta_\kappa^2}\left[ \sqrt{2} \beta_\kappa \sin
  \gamma_\kappa + \sqrt{\frac{1}{7}} \beta_\kappa^2 \chi_\kappa \sin
 (2\gamma_\kappa) \right].
\end{eqnarray}
\begin{center}
\begin{figure}[hbt]
\includegraphics[height=6cm,width=5cm]{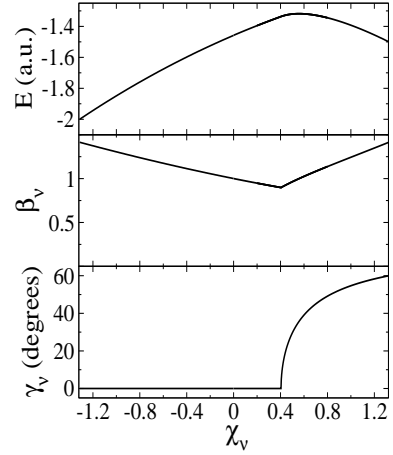}
\caption{Transition from $SU(3)$ to $SU^*(3)$: $x=0$,
$\chi_\pi=-\sqrt{7}/2$ and $\chi_\nu$ varies from $-\sqrt{7}/2$ to
$+\sqrt{7}/2$. In the panels are plotted the energy of the ground
state in arbitrary units, and the variation of the shape parameters
$\beta_\nu$ (dimensionless) and $\gamma_\nu$ (degrees).}
\label{FIG2}
\end{figure}
\end{center}
As a natural extension of the Casten triangle for IBM--1 \cite{Ca}, the
geometrical representation of the IBM--2 is a pyramid with the new
triaxial dynamical symmetry $SU^*(3)$ \cite{DB82} in the upper
vertex. Fig.~\ref{FIG1} shows a pictorial representation of the IBM--2
parameter space. Any point in this space is obtained with the following
transformation to polar coordinates (see Fig.~\ref{FIG1}).
\begin{equation}
\rho = 1-x ~;~ \theta = - \frac{\pi}{3}~\frac{\chi_\pi-\chi_\nu}
{\sqrt{7}} ~;~ \phi =  -
\frac{\pi}{3}~\frac{\chi_\pi+\chi_\nu}{\sqrt{7}} ~.
\end{equation}
We have explored the IBM-2 parameter space of Hamiltonian (\ref{HQ})
and
present here a selected set of calculations in order to establish
the IBM--2 phase diagram (a more detailed presentation will be given in
a forthcoming publication). We have not found traces of phase
transitions in the transition from $O(6)$ to $SU^*(3)$ in a parallel way
as the already known transition from $O(6)$ to $SU(3)$ in IBM--1. The $O(6)$
symmetry is in fact very unstable against small perturbations driving
the system out of the dynamical symmetry either to axial deformed
or to triaxial shapes depending on the interaction. The $O(6)$
symmetry itself has been proposed as a critical dynamical symmetry
\cite{Jol}.

In Fig.~\ref{FIG2} we show the transition $SU(3)\rightarrow SU^*(3)$
through the edge plotted in Fig.~1. Along this line $x=0$ and
$\chi_\pi=-\sqrt{7}/2$ are fixed. The relevant control parameter is
$\chi_\nu$ varying from $-\sqrt{7}/2$ (equal and aligned quadrupole
prolate shapes for protons and neutrons) to $\sqrt{7}/2$ (quadrupole
prolate shape for protons and quadrupole oblate shape for neutrons
with perpendicular axis of symmetry \cite{DB82}). In Fig.~\ref{FIG2}
we present the results for the ground state energy (in arbitrary
units) and the shape parameters ($\beta_\nu, \gamma_\nu$). The
resulting proton parameters are $\beta_\pi=\sqrt{2}$ and $ \gamma_\pi=0$ for
all values of the control parameter $\chi_\nu$. In the limit
$\chi_\nu=-\sqrt{7}/2$ we recover the results known from IBM-1:
$\beta_\nu=\sqrt{2}$ and $\gamma_\nu=0$.  In the opposite limit
$\chi_\nu=\sqrt{7}/2$ the results known from Ref.~\cite{DB82} are
obtained: $\beta_\nu=\sqrt{2}$ and $\gamma_\nu=60^{\circ}$.
Around $\chi_\nu=0.4035$ a clear shape phase transition is observed
changing the system from axial ($\chi_\nu<0.4035$) to triaxial
($\chi_\nu>0.4035$). We will call this point ``y''. Note that in this
phase transition the order
parameter is $\gamma_\nu$ that changes from $0$ in the symmetric phase
to a finite value in the non-symmetric phase \cite{Landau}. We have
minimised the energy following two inverse paths looking for possible
coexistence of minima and the corresponding spinodal and antispinodal
points. Both calculations give exactly the same
results. This means that spinodal, critical and antispinodal points
all converge to a single point \cite{IZ04}. Therefore, the transition
from $SU(3)$ to $SU^*(3)$ is second order. 
\begin{center}
\begin{figure}[hbt]
\includegraphics[height=6cm,width=5cm]{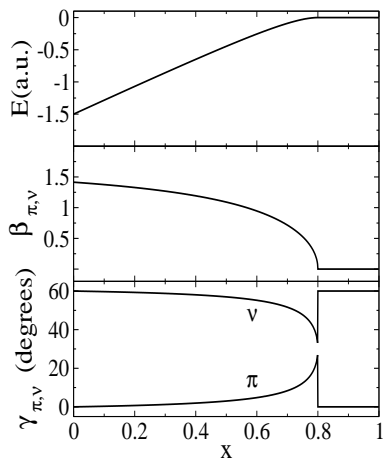}
\caption{Same as Fig.~\ref{FIG2} but for the
  transition from $U(5)$ to $SU^*(3)$:
  $\chi_\pi=-\chi_\nu=-\sqrt{7}/2$ and $x$ varies from $0$ (triaxial)
  to $1$ (spherical).}
\label{FIG3}
\end{figure}
\end{center}
Fig.~\ref{FIG3} shows the transition from $U(5) \rightarrow SU^*(3)$
through the corresponding edge in Fig.~\ref{FIG1}. Along this edge
$\chi_\pi=-\chi_\nu=-\sqrt{7}/2$ are fixed and the relevant control
parameter is $x$ changing from $1$ (spherical) to $0$ (triaxial). The
values of $\beta_\pi$ and $\beta_\nu$ are always equal at the energy
minimum, while $\gamma_\pi$ and $\gamma_\nu$ are symmetric with
respect to $\gamma = 30^{\rm \circ}$ axis. In the different panels,
the energy, and the values of $\beta$ and $\gamma$ for proton and neutron
shapes are presented. For $x=1$, $\beta_\pi=\beta_\nu=0$
implying a spherical shape. For $x=0$ we recover the $SU^*(3)$ case
with $\beta_\pi=\beta_\nu=\sqrt{2}$, and $\gamma_\pi=0$ and
$\gamma_\nu=60^{\rm \circ}$.
A phase transition at $x=0.8$ is observed. We will call this point
``x$^*$''. As in the preceding
case, we have performed two sets of calculations following inverse
paths  to determine the
order of the transition and again we have found no region of
coexistence, converging at the same place, spinodal, critical, and
antispinodal points.
Note that in
this case the order parameter is $\beta_\pi=\beta_\nu$, as well as
$\gamma_\pi=60-\gamma_\nu$.
\begin{center}
\begin{figure}[hbt]
\includegraphics[height=6cm,width=5cm]{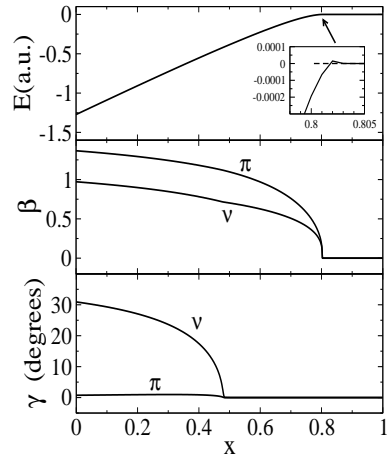}
\caption{Same as Fig. \protect\ref{FIG2} but for a generic
  transition from U(5) to a triaxial shape. The structure
  parameters are $\chi_\pi=-1.2$, $\chi_\nu=0.5$ and the control
  parameter $x$ varies from $0$ to $1$.}
\label{FIG4}
\end{figure}
\end{center}
In Fig.~\ref{FIG4}, we present the study of a generic transition from
$U(5)$ (spherical) to a triaxial shape through a trajectory within the
IBM--2 pyramid. In particular, we have selected the trajectory defined
by $\chi_\pi=-1.2$ and $\chi_\nu=0.5$, using $x$ as the control
parameter varying from $1$ to $0$. The ground state energy, and the
values for the $\beta$'s and the $\gamma$'s are plotted. Two phase
transitions are observed at different values of $x$. Starting from $x=1$
(spherical system), a first transition to axial deformed shape is
observed at $x\approx 0.8$. At this point, the values of $\beta_\pi$
and $\beta_\nu$ depart from zero but $\gamma_\pi$ and $\gamma_\nu$ are
zero indicating a deformed axial symmetry. $\beta_\pi$ and $\beta_\nu$
play the role of order parameters in this phase transition. For a
value of $x\approx 0.48$ a second phase transition is observed. The
values of $\beta_\pi$ and $\beta_\nu$ are different from zero in both
sides changing smoothly along the transition. The angular parameters $\gamma$
jump from zero to finite values indicating a transition from an
axial shape to a triaxial shape. Therefore $\gamma_\pi$ and $\gamma_\nu$
are the order parameters.
The different values for the shape
parameters for protons and neutrons are due to the selection of the
structure parameters $\chi_\pi$ and $\chi_\nu$ for this trajectory.
As in preceding
cases we have performed two sets of calculations following inverse
paths to determine the order of the phase transition.
The transition at $x\approx 0.8$ can be analysed looking at the behaviour of
the ground state energy (inset). Where the full line corresponds to a
forward calculation, starting at $x=0$, increasing $x$, and the dashed
line to a backward calculation, starting at $x=1$, decreasing
$x$. The inset shows that there are two minima competing, one
spherical and one deformed. If the system comes from the spherical
region it keeps spherical for a while even there is another deformed
minimum with slightly lower energy. On the other side, if the system
comes from the deformed region it keeps deformed (look at the small
peak in the full line in the inset at x=0.802) although another
spherical minimum have slightly
lower energy. This coexistence of deformed and spherical minima in a
small region around $x=0.8$ is the signature for a first order phase
transition.
The phase transition at $x\approx 0.48$ has been studied with forward
and backward calculations. We have not found any coexistence region.
The antispinodal, critical and spinodal points come together to a single
point as corresponds to a second order phase transition.
\begin{center}
\begin{figure}
\includegraphics[height=6cm,width=6cm]{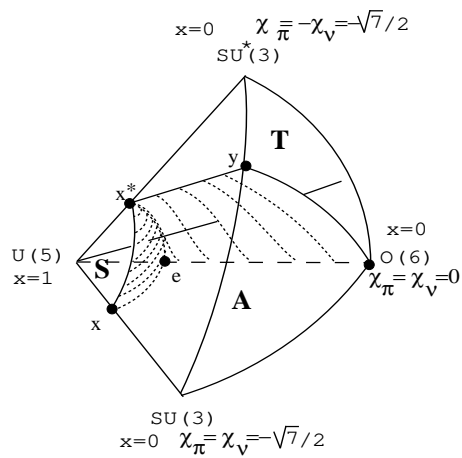}
\caption{Schematic phase diagram for IBM-2. S stands for spherical, A
  for axial and T for triaxial phases. The critical points
  ``x$^*$'' and ``y'' studied here and those already known for the
  IBM-1 phase diagram, ``x'', ``e'', and ``O(6)'', are marked with dots.
}
\label{FIG5}
\end{figure}
\end{center}
We have explored the parameter space of the IBM--2 Hamiltonian
(\ref{HQ}) in
Fig.~\ref{FIG1}.  The resulting phase diagram of the
proton--neutron IBM as described by the Hamiltonian (\ref{HQ}) is
depicted in Fig.~\ref{FIG5}. There are three well defined phases:
spherical, axially deformed (prolate in the schematic presentation
of Fig.~\ref{FIG5}) and triaxial.
The critical surface separating spherical and axially deformed shapes
(e--x$^*$--x--e) is first order, while the surface separating axially
deformed and triaxial shapes (e--O(6)--y--x$^*$--e) is second order,
including the common line between both surfaces (e--x$^*$).
We have checked that in all the cases discussed in which the
transition is second order the behaviour 
of the corresponding order parameter near
the critical point is consistent with a critical exponent $1/2$ as
given by the Landau theory \cite{Landau}.
We would like to stress that the critical
surface separating spherical and axially deformed nuclei is almost
a sphere with a radius equal to $\rho=0.2$ and centered in $U(5)$.
The straight line plotted inside
the figure gives an idea of the trajectory followed by the
transition discussed in Fig.~\ref{FIG4}. We would like to emphasise that we
have found a new critical point (x$^*$) at the phase transition
changing directly from
spherical to triaxial shapes. We are currently studying the
spectroscopic properties of this critical point. The
results will be presented elsewhere. Finally, this scheme of
analysis can be easily extended to positive values of $\chi_\pi$
to obtain the dynamical symmetry limits $\overline{SU(3)}$ and
$\overline{SU^*(3)}$.

We acknowledge discussions with F. Iachello and M. Caprio. This work
was supported in part by the Spanish DGI under project numbers
BFM2002-03315, BFM2000-1320-C02-02, BFM2003-05316-C02-02, and
FPA2003-05958.

\end{document}